\documentclass[conference]{IEEEtran}
\IEEEoverridecommandlockouts

\usepackage{cite}
\usepackage{amsmath,amssymb,amsfonts}
\usepackage{algorithmic}
\usepackage{graphicx}
\usepackage{textcomp}
\usepackage{xcolor}
\usepackage{booktabs}
\usepackage{multirow}
\usepackage{subcaption}
\usepackage[hyphens]{url}

\graphicspath{{./figures/}}

\def\BibTeX{{\rm B\kern-.05em{\sc i\kern-.025em b}\kern-.08em
    T\kern-.1667em\lower.7ex\hbox{E}\kern-.125emX}}

\usepackage[disable]{todonotes}

\begin{document}
\bstctlcite{IEEEexample:BSTcontrol}

\title{Automated Instruction Encoding Synthesis\\
       for Modern GPU ISA Compression}

\author{
\IEEEauthorblockN{Mingyuan Ma}
\IEEEauthorblockA{\textit{School of Integrated Circuits} \\
\textit{Tsinghua University} \\
Beijing, China \\
cnwhmmy@outlook.com}
\and
\IEEEauthorblockN{Hu He\textsuperscript{*}}
\IEEEauthorblockA{\textit{School of Integrated Circuits} \\
\textit{Tsinghua University} \\
Beijing, China \\
hehu@tsinghua.edu.cn}
\thanks{\textsuperscript{*}Corresponding author.}
\thanks{\copyright{} 2026 IEEE. Personal use of this material is permitted.
Permission from IEEE must be obtained for all other uses, in any current or
future media, including reprinting/republishing this material for advertising
or promotional purposes, creating new collective works, for resale or
redistribution to servers or lists, or reuse of any copyrighted component of
this work in other works.}
}

\maketitle

\begin{abstract}
Modern GPU kernels increasingly stress the instruction supply path,
while fixed instruction containers can leave substantial footprint
slack. This paper presents an automated encoding-synthesis framework
that treats instruction layout as a constrained slot-assignment
problem over a validated instruction-form field specification. The formulation
separates semantic field identity from physical bit positions and
supports tied, pinned, and free placement constraints, making it
applicable when recurring decoded fields are not frozen by a public
format contract. We instantiate the framework for NVIDIA SASS: raw public
encoding text is normalized into a machine-readable
specification, a SASS disassembler is validated against
\texttt{nvdisasm} on 3.78M instructions, released as an
open benchmark, and CP-SAT synthesis is used
for fixed-length and variable-length encodings. On 142 Blackwell
kernel inputs, variable-length synthesis reduces instruction footprint
by 33\%, with comparable reductions after re-synthesis on Ampere and
Hopper; fixed-length synthesis on the same specification reduces
decoder area by 16\% against a decoder generated from the
NVIDIA-observed 128-bit layout by the same generator and flow. Generated fetch/decode RTL meets 1.5\,GHz in
TSMC 22\,nm with a replicated area delta of 0.12\% of a GA100-class
die; a same-node SRAM comparison shows the footprint reduction
corresponds to about 9$\times$ this added logic in instruction-SRAM
bit-cell area.
\end{abstract}


\section{Introduction}
\label{sec:intro}

Modern GPU kernels increasingly stress the instruction supply path.
Fused operators, persistent kernels, warp-specialized pipelines, and
aggressive software pipelining place more static instructions inside
one kernel and can keep several disjoint code regions active on the same
streaming multiprocessor (SM). Public measurements and our microbenchmarks
(\S\ref{sec:bg-supply}) show instruction-supply thresholds in the
tens to hundreds of KiB. SASS, the native machine-level instruction
set of NVIDIA GPUs, encodes every instruction in 16 bytes, so these
thresholds correspond to only a few thousand static instructions.
Instruction footprint is therefore not only a binary-size property,
but also a frontend-capacity and fetch-bandwidth demand.

NVIDIA SASS in recent architectures (\texttt{sm\_70}
Volta through \texttt{sm\_100} Blackwell) uses an aligned fixed
128-bit container for every instruction. This uniform granularity is
paid by every dynamic instruction, even when the decoded operands
and modifiers require far fewer bits. Across our evaluated
\texttt{sm\_100} benchmark set, the frequency-weighted operand and
modifier fields occupy about 42~bits after opcode, predicate, and
shared control fields are separated. This gap exposes a code-density opportunity: a replacement encoding
can rearrange the same decoded fields while preserving instruction
semantics.

A replacement encoding must solve two coupled problems.
First, public SASS artifacts are not directly usable as an
encoding-synthesis input. Synthesis needs validated per-form field
structure:
which operands and modifiers each instruction form (one defined
instruction) carries, how
wide each field is, and which field values distinguish
similar forms. Second, a replacement encoding must place every field anew, a
decision no existing flow optimizes: fixed-format ISAs set
positions by hand in a small format catalogue, and automated
encoding synthesis optimizes opcode assignment while positions
follow the hand-written processor
description~\cite{nohl2003lisa,chattopadhyay2007power}. SASS exposes
what both paths miss: hundreds of operand and modifier fields recur
across 975 instruction forms, and their positions can be chosen
jointly (\S\ref{sec:bg-motivation}).

\begin{figure}[t]
\centering
\includegraphics[width=1.0\columnwidth]{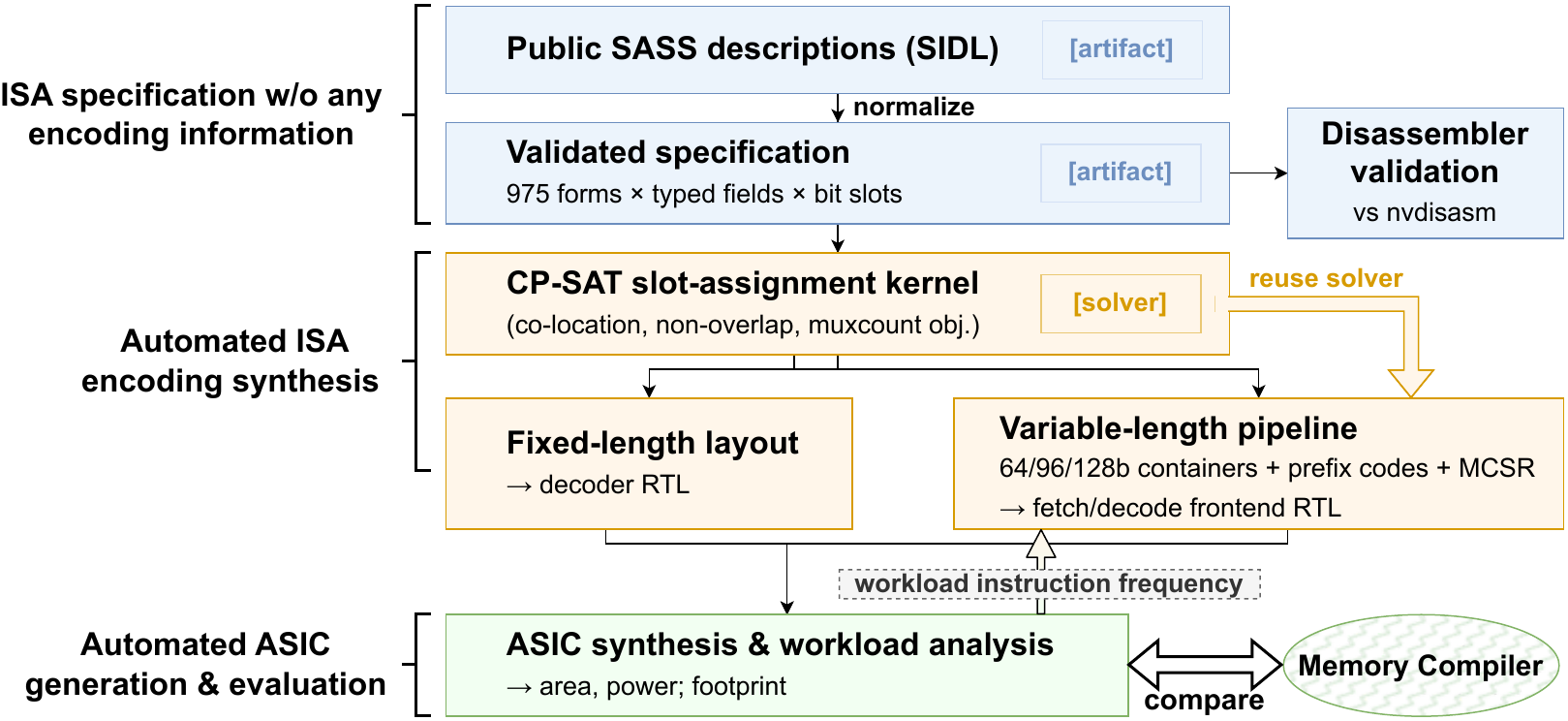}
\caption{End-to-end encoding-synthesis flow.}
\label{fig:flow}
\end{figure}

In this paper, we take a holistic optimization approach, deciding
all field positions jointly on a disassembler-validated benchmark
of the real SASS instruction set (Fig.~\ref{fig:flow}). First, we construct a
validated, machine-readable SASS benchmark from raw
instruction-description text publicly available from NVIDIA's
toolchain. Second, we formulate field placement as a constrained
slot-assignment problem and solve it with the CP-SAT
constraint-programming solver of OR-Tools for both
fixed-length and variable-length encodings. Finally, the encodings
are emitted as SystemVerilog fetch/decode frontends and
characterized through ASIC synthesis.

The paper makes four contributions:
\begin{itemize}
\item \textbf{A validated SASS specification and disassembler.}
  Raw public SASS encoding text is normalized into a typed,
  per-form field specification. The generated disassembler matches
  \texttt{nvdisasm} across 4 GPU generations, 3.78M instructions,
  and 30.0M field decodes. The toolchain is released at
  \url{https://github.com/reoLantern/nvsass-disassembler}.

\item \textbf{A CP-SAT formulation for SASS field layout.}
  Slot assignment combines form-local non-overlap, union-find field
  sharing, and a decoder mux objective. On the fixed 128-bit
  \texttt{sm\_100} container, synthesis reduces decoder area by 16\%
  and power by 21\% against
  the NVIDIA-observed layout.

\item \textbf{A workload-driven variable-length SASS encoding.}
  The 64/96/128-bit pipeline combines container assignment,
  prefix-free opcodes, cross-container field alignment, singleton
  absorption, and modifier-CSR offloading. It reduces Blackwell
  footprint by 33\%, with comparable Ampere and Hopper re-synthesis
  results.

\item \textbf{Generated RTL and ASIC characterization.}
  Auto-generated decoders and variable-length fetch/decode frontends
  meet 1.5\,GHz in TSMC 22\,nm; the replicated frontend-area delta is
  about 0.12\% of a GA100-class die, and the saved instruction-SRAM
  bit-cell area is about 9$\times$ this added logic at the same node.
\end{itemize}

\section{Background and Motivation}
\label{sec:background}

\subsection{Instruction-supply pressure in large GPU kernels}
\label{sec:bg-supply}

Recent GPU workload and kernel-optimization trends increase the
instruction footprint that on-chip instruction caches must hold.
Kernel fusion and persistent kernels are now standard practice for
AI workloads: fused attention~\cite{dao2022flashattention},
expert-parallel mixture-of-experts (MoE) kernels~\cite{deepep}, and
persistent megakernels~\cite{tilert} collapse multiple operator
phases into one kernel body, so a single launch carries a large
static instruction block. Aggressive unrolling and software
pipelining, used to hide memory latency, further enlarge the hot
code region by replicating steady-state instruction sequences. Warp
specialization adds pressure on an orthogonal axis: warps split into
producer, consumer, and epilogue roles, so each SM simultaneously
executes multiple disjoint instruction regions rather than a single
homogeneous code path.

\begin{table}[t]
\centering
\caption{NVIDIA instruction-cache capacities and level names as
reported by public microbenchmark studies, shown alongside documented
SM-local storage resources for scale. SMEM is the shared memory and RF
the register file.}
\label{tab:nvidia-icache}
\small
\begin{tabular}{lcccc}
\toprule
GPU & L0 I\$\textsuperscript{$\dagger$} & L1 I\$\textsuperscript{$\ddagger$} & SMEM & RF \\
\midrule
GV100~\cite{jia2018volta}
  & 48\,KiB & 128\,KiB & 96\,KiB & 256\,KiB \\
TU104~\cite{jia2019turing}
  & ${\approx}$64\,KiB & ${\approx}$46\,KiB & 64\,KiB & 256\,KiB \\
GB202~\cite{lam2025blackwell}
  & ${\approx}$128\,KiB & ${\approx}$128\,KiB & 228\,KiB & 256\,KiB \\
\bottomrule
\multicolumn{5}{l}{\scriptsize $\dagger$\,Reported per SM partition; shown as the
per-SM aggregate over 4 partitions.} \\
\multicolumn{5}{l}{\scriptsize $\ddagger$\,Named differently across these studies: an L1
instruction cache, or a} \\
\multicolumn{5}{l}{\scriptsize level merged with the L1.5 constant cache.}
\end{tabular}
\end{table}

\begin{figure}[t]
\centering
\includegraphics[width=0.85\columnwidth]{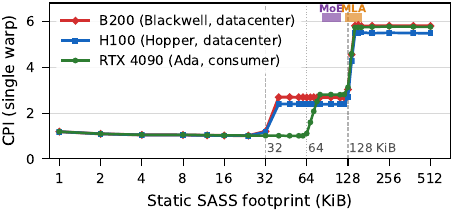}
\caption{Instruction-supply thresholds from a single-warp FFMA
hot-loop sweep on three NVIDIA GPUs. The vertical axis is median
cycles per \texttt{FFMA}; the horizontal axis is the static
footprint of the loop body. Top bands mark static TileRT
kernel footprint ranges; the sweep isolates frontend thresholds
rather than modeling those kernels directly.}
\label{fig:icache-microbench}
\end{figure}

Instruction storage is already large enough to consume meaningful
SM-local SRAM area, yet small enough to constrain these kernels.
Public microbenchmark studies place per-SM instruction storage in the
tens to hundreds of KiB and a further cache level above it
(Table~\ref{tab:nvidia-icache}), the same order of magnitude as
shared memory and the register file; enlarging this storage competes
with other SM-local memories for area. This capacity is still modest
in instruction terms: with 16-byte SASS instructions, 32, 64, and
128\,KiB hold only 2{,}048, 4{,}096, and 8{,}192 instructions before conflicts
and replacement reduce effective capacity. A hot-loop sweep isolates
this limit. One resident warp executes an
unrolled fused multiply-add (\texttt{FFMA}) loop with 64 independent
accumulator chains, whose static footprint is swept from 1 to
512\,KiB; each point reports the median cycles per \texttt{FFMA} (CPI)
over 30 runs. No data dependence and no co-resident warp compete for
issue, so the CPI steps mark instruction-supply limits. B200 and H100 first
transition at 32\,KiB, while RTX~4090 remains flat to 64\,KiB; all
three transition again near 128\,KiB
(Fig.~\ref{fig:icache-microbench}). NVIDIA reports the same H100
trend: larger workloads raise No-Instruction stalls and
instruction-cache misses, and reducing hot footprint reduces
both~\cite{nvidia2024icache}. TileRT~\cite{tilert}, a
state-of-the-art open-source LLM inference runtime, reaches this
scale in precompiled Blackwell megakernels: MoE kernels occupy
82--114\,KiB and multi-head latent attention (MLA) kernels occupy
121--163\,KiB of SASS text. Static footprint upper-bounds the hot
working set, but these ranges already overlap the measured frontend
thresholds.

\subsection{Cross-form field placement remains unoptimized}
\label{sec:bg-motivation}

Instruction-supply pressure motivates encoding synthesis rather
than larger instruction storage, but existing synthesis mostly optimizes
opcode assignment~\cite{nohl2003lisa,chattopadhyay2007power}, not
the placement of operand and modifier fields: the layout of a
replacement encoding is still decided by hand. In the \texttt{sm\_100} field
specification, instruction forms carry a median of 3 semantic
modifiers, close to the median 4 register/predicate operands, across
388 different modifier types. One rounding-mode modifier alone recurs
across 15 mnemonics, from FFMA to the conversions F2F and I2F. These
recurring operands and modifiers can be
placed globally once semantic field identity is separated from physical
bit positions: fields that co-occur in one form must be disjoint, while
fields that never co-occur may share a position. A layout with fewer
positions per decoded output reduces extraction muxing.

At 975 instruction forms, this cross-form placement choice is the
core of the synthesis problem. In encoding synthesis from hierarchical processor
models, a field
keeps one position inside the group that declares it, so positions
follow the designer's grouping; the line continues in modern
architecture-description-language (ADL) toolchains (\S\ref{sec:rw-encoding}).

\subsection{A benchmark gap for modern GPU encodings}
\label{sec:bg-spec}

NVIDIA GPUs dominate modern accelerated computing and are a natural
target for ISA-encoding research, yet no open, machine-readable SASS
benchmark with per-form field structure exists to study; we release a
validated specification toolchain and benchmark to close this gap.
Disassembly and opcode tables expose rendered instructions, not the
per-form field structure that synthesis needs. Community assemblers and
microbenchmarking studies cover useful subsets of the
encoding~\cite{cuassembler,jia2018volta,jia2019turing}, but operand
kinds, modifier domains, and architecture changes remain case-specific.

DocumentSASS~\cite{documentsass} supplies the public starting point by
capturing the Shader Instruction Description Language (SIDL) text that
\texttt{nvdisasm} expands while processing a cubin. Raw SIDL is text
written for the disassembler, not a synthesis-ready specification: its
encoding slots, table calls, and conditional rewrites must be
cross-referenced before each field occurrence has a precise role,
domain, and bit position.

The benchmark is constructed in \S\ref{sec:foundations};
\S\ref{sec:fixed} and \S\ref{sec:varlen} formulate fixed- and
variable-length placement; \S\ref{sec:eval} evaluates footprint and
RTL cost.

\section{Validated SASS Encoding Benchmark}
\label{sec:foundations}

We construct validated SASS specifications as a benchmark: the encoding is recovered from
public DocumentSASS text and checked field-by-field against
\texttt{nvdisasm} at scale. The benchmark normalizes raw SIDL into a
typed per-form specification, representing each instruction form as a
record of its operand and modifier fields, constant discriminators, and
table-encoded fields.

\subsection{Cross-referenced SIDL field interpretation}
\label{sec:found-sidl}

The SIDL-to-spec translation resolves field meaning from cross-file
references rather than individual \texttt{ENCODING} tokens. A
field that appears in assembly text is introduced in \texttt{FORMAT}
as a variable bound to a type and, often, a default value. The
top-level type definition supplies the legal labels and numeric
values for that variable. The \texttt{ENCODING} section places the
variable, a constant, an opcode value, or a table lookup into a
physical bit slot. The structured specification keeps these roles
separate: semantic field, value domain, discriminator status, and
encoded location are distinct attributes
(Table~\ref{tab:sidl-mapping}).
For example, \texttt{FFMA} binds printable modifiers to semantic
variables \texttt{fmz}, \texttt{rnd}, and \texttt{sat}, while
\texttt{ENCODING} maps them to physical slots such as
\texttt{fmz}, \texttt{stride}, and \texttt{NaN}. Thus \texttt{.RZ}
is a value of \texttt{rnd} even when its bits occupy the slot named
\texttt{stride}.

\begin{table}[t]
\centering
\caption{SIDL cross-references normalized into the structured SASS
specification.}
\label{tab:sidl-mapping}
\footnotesize
\begin{tabular}{@{}ll@{}}
\toprule
Raw SIDL relation & Structured-spec object \\
\midrule
\texttt{FORMAT\,TYPE(d):x} + top-level \texttt{TYPE}
  & typed field \texttt{x}, domain, default \\
\texttt{ENCODING\,slot\,=\,x}
  & decoded field \texttt{x} in physical slot \\
\texttt{ENCODING\,slot\,=\,Opcode}/const.
  & opcode / constant discriminator \\
\texttt{ENCODING\,slot\,=\,TABLES\_t(\ldots)}
  & atomic table-encoded field \\
\texttt{CLASS} / \texttt{ALTERNATE CLASS}
  & normalized instruction form \\
\bottomrule
\end{tabular}
\end{table}

This translation yields 975 instruction forms and
hundreds of distinct field names for \texttt{sm\_100}. Each
form records the fields required to decode and
re-encode it together with the constants that distinguish it
from other forms sharing the same opcode or assembly mnemonic family (e.g., all \texttt{FFMA} variants). The
same translation runs unchanged on the SIDL tables for 4
architecture generations (\texttt{sm\_80}--\texttt{sm\_100}).

\subsection{Field-level validation against \texttt{nvdisasm}}
\label{sec:found-disasm}

We validate the interpretation with two independently implemented
paths over the same instruction stream. The binary path matches each
128-bit word against the structured specification and extracts the
selected instruction form and its field values. The text path
parses the \texttt{nvdisasm} rendering using the SIDL \texttt{FORMAT}
string and reconstructs the fields visible in the assembly text. Agreement is checked at three levels: whether the binary path
selects a unique instruction form (T0), whether the normalized
mnemonic matches (T1), and whether the field values decoded on both
sides agree (T2). T2 is the critical level: it catches cases where both
paths agree on instruction identity but disagree on individual field
values.

We run this validation across 866 benchmarks spanning the 4
architecture generations, drawn from public GPU benchmark suites
(Rodinia, Polybench-GPU, Parboil, SHOC, microbench collections, and
CUTLASS / DeepGEMM kernels) and compiled per architecture with
CUDA~13.1. Table~\ref{tab:disasm-validation} reports the result.
Across all 4 generations, every instruction resolves to a unique
form, matches \texttt{nvdisasm} on the mnemonic, and agrees on every
jointly decoded field value, confirming 30.0 million field decodes.

The last two columns report which part of each specification the
stream exercises. Compiled CUDA code reaches 354 of the 975
\texttt{sm\_100} forms and 147 of its 230 mnemonics; many forms it
never selects belong to graphics and fixed-function units that
\texttt{nvcc} does not emit, such as attribute access
(\texttt{AL2P}, \texttt{ALD}, \texttt{AST}) and texture-cache
control (\texttt{CCTLT}). Validation therefore covers the fields
this corpus exercises, while the
synthesis of \S\ref{sec:fixed} and \S\ref{sec:varlen} places the
fields of all 975 forms, exercised or not.

\begin{table*}[t]
\centering
\caption{Disassembler validation: 3.78M instructions across 4 GPU generations match \texttt{nvdisasm} on the mnemonic and on every jointly decoded field. The last two columns give the instruction forms and mnemonics the corpus exercises out of those the specification defines.}
\label{tab:disasm-validation}
\small
\begin{tabular}{lrrrrrrr}
\toprule
Architecture & Benchmarks & Instructions & Form+mnem pass & Fields checked & Field agree & Forms used & Mnem.\ used \\
\midrule
\texttt{sm\_80}   (Ampere)    & 217 & 1{,}074{,}312 & 100.000\% &  7{,}660{,}832 & 100.000\% & 287 / 962  & 103 / 164 \\
\texttt{sm\_89}   (Ada)       & 220 &    763{,}368 & 100.000\% &  6{,}035{,}290 & 100.000\% & 294 / 1{,}048 & 107 / 175 \\
\texttt{sm\_90}   (Hopper)    & 212 &    619{,}544 & 100.000\% &  5{,}062{,}986 & 100.000\% & 292 / 1{,}167 & 125 / 206 \\
\texttt{sm\_100} (Blackwell) & 217 & 1{,}320{,}040 & 100.000\% & 11{,}194{,}766 & 100.000\% & 354 / 975  & 147 / 230 \\
\bottomrule
\end{tabular}
\end{table*}

\subsection{Synthesis input as a per-form field projection}
\label{sec:found-input}

Synthesis uses a projection of the full specification.
Let $\mathcal{L}$ denote the set of instruction forms. A \emph{field
occurrence} is one instance of a decoded field in one form. For each
form~$\ell \in \mathcal{L}$, the specification yields a set~$O_\ell$
of field occurrences, each carrying a tuple
$(n, w, k, D)$: semantic field name~$n$, encoded width~$w$, field
kind~$k$ (operand, operand modifier, standalone modifier, or
table-encoded field), and legal value domain~$D$. Dense register and
immediate domains are represented by type and width rather than by
enumerating all values.

Scheduling control, predicate guards, opcode bits, and constant
discriminators occupy fixed container regions and are excluded
from~$O_\ell$. Table calls remain atomic fields with their encoded
widths and output domains.

\section{Fixed-Length Encoding Synthesis}
\label{sec:fixed}

\begin{figure}[t]
\centering
\includegraphics[width=0.85\columnwidth]{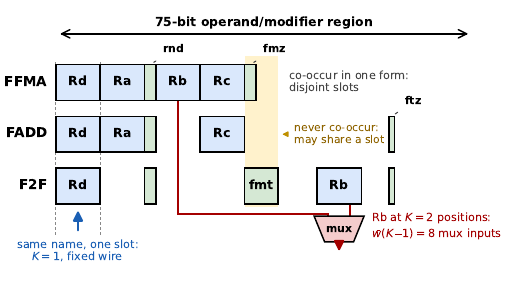}
\caption{Cross-form slot assignment over three \texttt{sm\_100}
forms in the 75-bit operand/modifier region of
\S\ref{sec:fixed-model}.}
\label{fig:placement-schematic}
\end{figure}

One slot-assignment model underlies both encodings synthesized in
this paper, the fixed-length encoding of this section and the
variable-length encoding of \S\ref{sec:varlen}
(Fig.~\ref{fig:placement-schematic}). Each recurring field group
picks one slot, free by default unless designer constraints tie or
pin it; co-occurring fields of
a form occupy disjoint slots, never-co-occurring fields may stack
(\texttt{fmz}/\texttt{fmt}), and a field name extracted from two
positions, like \texttt{Rb}, pays mux inputs for the extra one.
This section instantiates the model in the fixed 128-bit container
with a decoder-muxing cost; \S\ref{sec:varlen} adds per-form
container classes with prefix-code budgets and switches the cost to
frequency-weighted footprint.

\subsection{Field-region placement model}
\label{sec:fixed-model}

Given the occurrence sets~$O_\ell$ from \S\ref{sec:found-input}, the
fixed-length formulation places operand and modifier fields inside the
128-bit container. The container allocates 22~bits to scheduling
control, 4 to the predicate guard, and 13 to the opcode; 14~bits are
unused in the reference encoding. These regions remain at their
reference positions, leaving a contiguous $R = 75$-bit operand/modifier region
$B = \{0, \ldots, R{-}1\}$ for operand and modifier placement.

Our model uses a pre-enumerated slot pool rather than continuous
integer bit-position variables. For a field of width~$w$, a
contiguous slot starting at bit~$b$ is the set
\[
  s(b,w) = \{b,\; b{+}1,\; \ldots,\; b{+}w{-}1\}\,.
\]
The width-$w$ slot pool is
\[
  \mathcal{S}_w =
  \bigl\{s(b,w) : b \in B,\; b{+}w \le R\bigr\}\,.
\]
All results use contiguous slots; the formulation admits any
finite slot pool.

The slot model exposes placement constraints at occurrence
granularity. A \emph{co-location constraint} states that two
equal-width field occurrences must choose the same slot; such
constraints may come from a global field name, a designer-defined
instruction class, an operand role, or an individual occurrence pair.
These constraints are not tied to a fixed ADL hierarchy: nested or
overlapping classes are compiled into pairwise relations; a
union-find equivalence pass computes their transitive closure before
CP-SAT variable creation, so many occurrence-level decisions
collapse into one shared slot decision. The closure partitions
$O = \bigcup_\ell O_\ell$ into \emph{assignment
groups}~$\mathcal{G}$, each sharing width~$w_g$ and slot choice. A
group may contain different field names when they fill the same
decoder role or are intentionally unified by a placement constraint
(e.g., \texttt{Rd} and \texttt{URd} as destination registers). The
solver assigns each group to exactly one slot:
\[
  x_{g,s}\in\{0,1\},
  \qquad
  \forall g\in\mathcal{G}:\quad
  \sum_{s\in\mathcal{S}_{w_g}} x_{g,s}=1 .
\]
The packing constraint is form-local bit non-overlap: for every form
and every bit in the placement region,
\[
  \forall \ell\in\mathcal{L},\;\forall b\in B:\quad
  \sum_{\substack{g\in\mathcal{G}:\\ g\cap O_\ell\ne\emptyset}}
  \sum_{\substack{s\in\mathcal{S}_{w_g}:\\ b\in s}}
  x_{g,s} \le 1 .
\]

\subsection{Decoder-cost objective}
\label{sec:fixed-cost}

Let $N$ denote the set of distinct field names across all
forms. For a given assignment~$x$, let $P_n(x)$ be the set of
low-bit positions from which field name~$n$ may be extracted:
\[
  P_n(x) =
  \bigl\{\operatorname{lo}(s) :
    x_{g,s}=1,\; n \in \mathrm{names}(g)\bigr\}\,,
\]
where $\mathrm{names}(g)$ collects the field names of all occurrences
in~$g$. The number of extraction positions is
$K_n(x) = |P_n(x)|$. When $K_n = 1$, the decoder wires field~$n$
from a fixed bit slice; when $K_n > 1$, it selects among $K_n$
positions once the form is known
(Fig.~\ref{fig:placement-schematic}).

A $K$-way selection can be implemented as a tree of $K{-}1$
two-input muxes per output bit. With $\bar{w}_n$ denoting the
maximum output width emitted for name~$n$, the mux-cost proxy
(the \emph{muxcount}) therefore counts the two-input muxes across all
decoder outputs:
\[
  \widehat{M}(x) =
  \sum_{n\in N} \bar{w}_n\,\bigl(K_n(x)-1\bigr)\,.
\]
It is an estimate, not a
technology-mapped area formula.

The complete fixed-length synthesis problem is
\[
\begin{aligned}
  \min_{x}\quad
    & \widehat{M}(x)
      = \sum_{n\in N} \bar{w}_n\bigl(K_n(x)-1\bigr) \\
  \textrm{s.t.}\quad
    & \sum_{s\in\mathcal{S}_{w_g}} x_{g,s}=1
      && \forall\, g\in\mathcal{G},\\
    & \sum_{\substack{g\in\mathcal{G}:\, g\cap O_\ell\ne\emptyset}}
      \;\sum_{\substack{s\in\mathcal{S}_{w_g}:\, b\in s}}
      x_{g,s} \le 1
      && \forall\, \ell\in\mathcal{L},\; b\in B,\\
    & K_n(x) \le c_n
      && \forall\, n\in N_{\mathrm{cap}}\,.
\end{aligned}
\]
The set $N_{\mathrm{cap}}$ contains field names with explicit
dispersion bounds: $c_n = 1$ pins a name to a single position;
$c_n > 1$ limits but does not fix its coordinates.

\subsection{Constraint instantiation and solver instance}
\label{sec:fixed-priors}

The fixed-length experiment instantiates this interface with three
constraint sources.

\paragraph{Operand-role ties}
Within a mnemonic, different forms may spell the same operand role
with different field names. An FMA-like mnemonic with
\texttt{RRR}, \texttt{RRI}, \texttt{RIR}, \texttt{RRU}, and
\texttt{RUR} forms may use \texttt{Rb}, \texttt{Rc}, \texttt{URb},
or \texttt{URc} for the surviving source register. The operand
role, not the local field name, determines the equality group. Operand
modifiers attached to a merged base operand follow the same group when
the modifier kind exists on both sides. A second pass chains register
groups that occupy the same physical position in the reference NVIDIA
encoding across mnemonics, collapsing high-frequency register names
to $K_n = 1$ globally.

\paragraph{Discriminator ties}
Some forms have the same opcode and fixed constants in the reference
encoding, so a variable field with non-overlapping legal values
selects the form. The fixed-length experiment preserves this shared
identity rather than assigning new opcodes. The replacement layout
therefore ties each such distinguishing field to one position across
the affected forms, keeping the same form-selection test available.
This rule creates 138 shared-identity groups and 457 pairwise
distinguishing-field tie constraints.

\paragraph{Name-dispersion caps}
Audited field names whose values feed common decoder outputs receive
global pins or small caps on $K_n$. For example, setting
$K_{\texttt{Rd}}{=}1$ forces every destination-register occurrence to
use one extraction position; small caps allow limited dispersion for
less regular fields. They express designer preferences without fixing
coordinates: the solver still chooses legal slots under form-local
non-overlap and the mux proxy.

The Boolean model is instantiated with OR-Tools CP-SAT. The
\texttt{sm\_100} fixed-layout instance contains 975 instruction
forms, 7453 field occurrences after fixed-region exclusions, 2450
assignment groups, and 1404 candidate slots. The resulting model has 178{,}841 assignment
variables and 72{,}029 form-level non-overlap constraints.

\section{Variable-Length Encoding Synthesis}
\label{sec:varlen}

Variable-length synthesis runs in three stages: container classes
and two bit-reduction mechanisms, singleton absorption and
modifier control-and-status register (MCSR) offloading, set each
form's field budget
(\S\ref{sec:varlen-payload}); a CP-SAT assignment picks each form's
container class and identity-code length under a prefix-free budget
(\S\ref{sec:varlen-assign}); the slot-assignment model of
\S\ref{sec:fixed} then places fields in the remaining bits
(\S\ref{sec:varlen-fields}).

\subsection{Field budgets and container classes}
\label{sec:varlen-payload}

Variable-length encoding couples field placement to a container
choice for each instruction form. Each form selects one class
$c \in \mathcal{C}=\{64,96,128\}$ bits, and the objective becomes the
frequency-weighted total encoding length.

\begin{figure}[t]
\centering
\includegraphics[width=0.9\columnwidth]{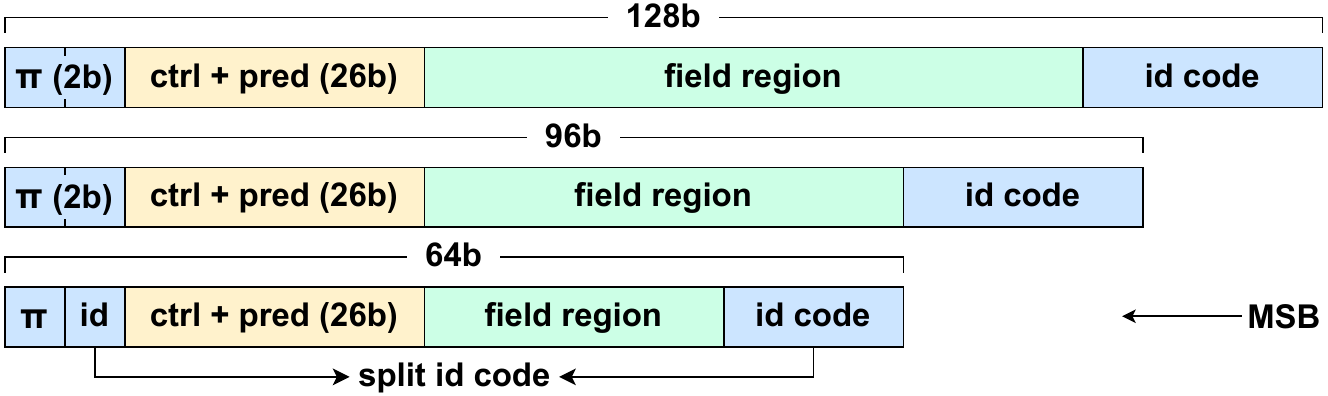}
\caption{Container bit layout for variable-length classes.}
\label{fig:container-layout}
\end{figure}

Each container class~$c$ reserves $\pi_c$~bits for a length prefix
($\pi_{64} = 1$; $\pi_{96} = \pi_{128} = 2$) and 26~bits for
scheduling control and predicate guard. The remaining budget for the
per-form identity code (the prefix-free opcode bits that identify each
form within its container class) and operand/modifier fields is
\[
  B_c = c - 26 - \pi_c\,,
\]
giving $B_{64} = 37$, $B_{96} = 68$, $B_{128} = 100$.
For 96- and 128-bit containers, bits~0--1 hold the prefix, bits~2--27
hold control and predicate fields, field bits start at bit~28, and the
identity code occupies the MSB end. The 64-bit container uses bit~0 for
the prefix and bit~1 for one identity-code bit, keeping the field bits
aligned at bit~28 (Fig.~\ref{fig:container-layout}).
After container assignment, branch targets denote container starts and
branch displacements are recomputed over the 32-bit-word-aligned
synthesized stream.

The variable-length formulation consumes two workload-derived
quantities. Binary matching against the validated specification
(\S\ref{sec:found-disasm}) gives an instruction-stream count~$f_\ell$ for each form. Modifier
traces along the instruction stream give a change
rate for each \emph{modifier class}, identified by its semantic
name~$n_o$ and enumeration type~$t_o$: the fraction of consecutive
occurrences of that class whose decoded value differs from
the previous occurrence.
These frequencies and change rates are treated as inputs to the
synthesis problem; the benchmark set and resulting statistics are
reported in \S\ref{sec:eval}.

The variable-length design assigns each form its own prefix-free
identity code rather than preserving NVIDIA's shared opcode and
constant-discriminator structure. A modifier whose value domain has
size one in a given form is therefore implied by the form identity and
need not occupy container bits. Let $A_\ell$ denote the set of such
\emph{singleton} occurrences:
\[
  A_\ell = \bigl\{o \in O_\ell :
    k_o = \text{modifier},\; |D_o| = 1\bigr\}\,.
\]
A second reduction targets slow-changing modifiers. A
\emph{modifier control-and-status register (MCSR)} file holds
modifier values that persist across instructions; a dedicated
\texttt{MCSR-SET} instruction updates the register when the value
changes. A modifier class whose change rate falls below a threshold
is eligible for MCSR offloading; offloaded values are supplied by
architectural state rather than encoded in every instruction. Let
$S_{\mathrm{MCSR}}$ denote the set of selected (name, type) pairs
and
\[
  H_\ell = \bigl\{o \in O_\ell :
    (n_o, t_o) \in S_{\mathrm{MCSR}}\bigr\}\,.
\]
The effective field-bit budget for form~$\ell$ is then
\[
  P_\ell = \sum_{o \in O_\ell \setminus
    (A_\ell \cup H_\ell)} w_o\,.
\]
The MCSR file is per-warp architectural state: each warp slot carries
its own copy, since warps resident in a subcore interleave and each
advances its own instruction stream. For \texttt{sm\_100} the selected
set holds 8 modifier classes, 15 bits per warp, so the 16 warp slots
of a GA100-class subcore~\cite{nvidia-a100-whitepaper} store 30 bytes
of MCSR state, saved and restored with the rest of the warp context.
MCSR offloading is not free: whenever an offloaded modifier changes value
along the instruction stream, an \texttt{MCSR-SET} instruction
must be inserted. The code-size evaluation in \S\ref{sec:eval}
accounts for these insertions by scanning instruction streams and adding
the \texttt{MCSR-SET} container cost at each transition.

\subsection{Length-class and prefix-code assignment}
\label{sec:varlen-assign}

The rate decision assigns each form a container class and an
identity-code length $q \in \{1, \ldots, q_{\max}\}$. Let
$y_{\ell,c,q} \in \{0,1\}$ indicate that form~$\ell$ selects
container~$c \in \mathcal{C}$ with code length~$q$; variables are
created only when $P_\ell + q \le B_c$. The assignment problem is
\[
\begin{aligned}
  \min_{y}\quad
    & \sum_{\ell,c,q} f_\ell \cdot c \cdot y_{\ell,c,q} \\
  \textrm{s.t.}\quad
    & \sum_{c,q} y_{\ell,c,q} = 1
      && \forall\,\ell \in \mathcal{L},\\
    & \sum_{\ell,q} y_{\ell,c,q}\, 2^{-q} \le 1
      && \forall\,c \in \mathcal{C}\,.
\end{aligned}
\]
The first constraint assigns each form to exactly one
(class, code-length) pair. The second is the standard Kraft condition
for a prefix-free codebook within each container class. The objective
minimizes total frequency-weighted encoding length while leaving
enough per-container capacity for the fields placed in
\S\ref{sec:varlen-fields}. The Boolean model is solved with CP-SAT to
determine the minimum-cost container assignment. Within that assignment, code
lengths are tightened to the shortest feasible prefix-free values
before field placement. Codewords are then assigned by a standard
prefix-code construction.

\subsection{Field placement in variable containers}
\label{sec:varlen-fields}

Once each form has a container class~$c_\ell$ and identity-code
length~$q_\ell$, the field-region size for form~$\ell$ is
\[
  R_\ell = B_{c_\ell} - q_\ell\,.
\]
The field-placement model of \S\ref{sec:fixed-model} applies with one
change: the per-form slot budget is~$R_\ell$ rather than a
uniform~$R = 75$. A group~$g$ that spans multiple forms is
restricted to slots that fit every member:
\[
  b + w_g \le \min_{\ell:\, g \cap O_\ell \ne \emptyset} R_\ell\,.
\]
The mux-cost objective~$\widehat{M}$ and the operand-role and
cross-mnemonic ties of \S\ref{sec:fixed-priors} carry over.
Discriminator ties are no longer a correctness requirement
(per-form identity codes make shared-identity separation redundant)
but are retained as a decoder-quality heuristic.
Because the operand/modifier region starts at absolute container bit~28 in all
three classes, a field at relative
position~$b$ occupies the same hardware wire in every container. The
model exploits this by pinning shared register names
$\{\texttt{Rd}, \texttt{URd}, \texttt{Ra}, \texttt{URa}\}$ to
identical positions across all classes, collapsing register-field
extraction to a single hard-wire slice with no cross-class
multiplexer.

Four named \emph{encoding variants} configure the field-bit reduction:
\textit{pure} applies only variable containers and prefix codes;
\textit{+singleton} absorbs form-implied singleton modifiers;
\textit{+MCSR} adds MCSR offloading;
\textit{full} enables both mechanisms.
Each variant is solved end to end; their contributions are separated
in \S\ref{sec:eval-decoder} and \S\ref{sec:eval-footprint}.

\section{Experimental Evaluation}
\label{sec:eval}

\subsection{Experimental setup and RTL model}
\label{sec:eval-setup}

\begin{figure*}[t]
\centering
\includegraphics[width=\textwidth]{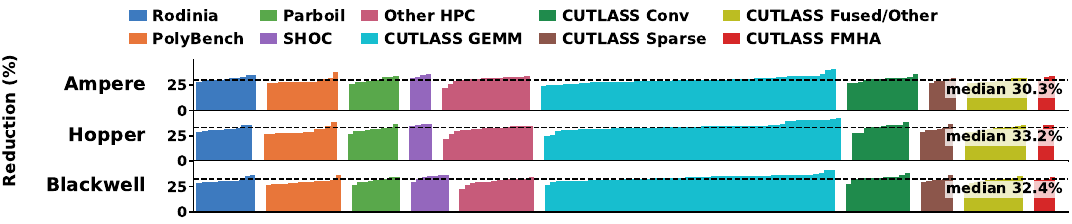}
\caption{Per-input footprint reduction after full re-synthesis from
  each architecture's own specification.}
\label{fig:per-benchmark}
\end{figure*}

Specification validation uses the benchmark set in
Table~\ref{tab:disasm-validation}. Encoding experiments use 142
nonempty Blackwell SASS inputs compiled with
CUDA~13.1 from Rodinia, PolyBench, Parboil, SHOC, other HPC
benchmarks, and CUTLASS examples for GEMM, convolution, sparse GEMM,
fused multi-head attention, and fused kernels. Binary
matching against the validated specification decodes 1{,}107{,}272
instruction instances; these matches provide per-form frequencies
for variable-length assignment. MCSR variants additionally scan each
matched instruction stream in program order to estimate modifier
value-change rates. The cross-generation footprint study
(Fig.~\ref{fig:per-benchmark}) re-runs the full pipeline on matching
Ampere (\texttt{sm\_80}, 139 inputs) and Hopper (\texttt{sm\_90}, 134
inputs) specifications. Each architecture uses its own instruction
frequencies and synthesized codebook, with no transfer across
generations. All constraint models use Google OR-Tools
CP-SAT on 16 threads.

\begin{figure*}[t]
\centering
\includegraphics[width=0.9\textwidth]{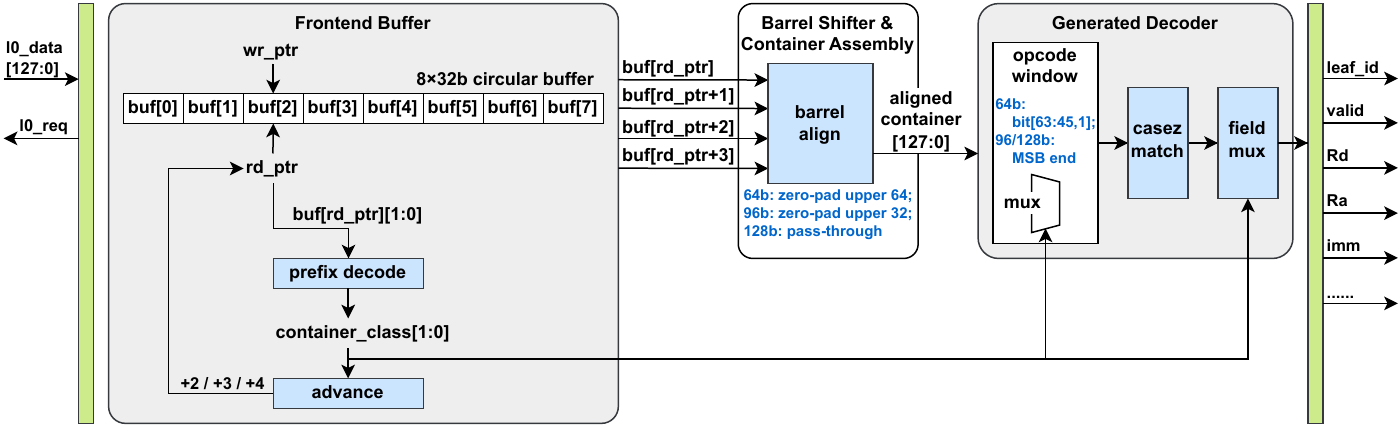}
\caption{Variable-length fetch-decode frontend.}
\label{fig:frontend-varlen}
\end{figure*}

Hardware evaluation uses subcore-level fixed-length and
variable-length fetch-decode RTL models.
The fixed-length path advances by 4 32-bit words per instruction.
The variable-length path reads prefix bits at a container start, waits
until the selected 2/3/4-word container is buffered, aligns it from the
circular word buffer, and advances by that length
(Fig.~\ref{fig:frontend-varlen}).

All decoders are auto-generated from the structured specification and
output the decoded SASS field vector: register addresses, immediates,
predicates, modifiers, and table-encoded values. Execution-control
generation for a proprietary GPU datapath is outside this
field-extraction scope. The fixed-length baseline uses NVIDIA's
observed 128-bit layout and opcode structure; the variable-length path
uses the synthesized containers and prefix codes. Both share the same L1-side interface,
pipeline boundary, Synopsys Design Compiler flow, CLN22ULP library,
1.5\,GHz target, and \texttt{compile\_ultra} settings, isolating the
encoding choice.

\subsection{Decoder synthesis and slot-assignment quality}
\label{sec:eval-decoder}

\begin{table}[t]
\centering
\caption{Fixed-length (128-bit) decoder synthesis. Both designs use the
same 975-form specification and ASIC flow (TSMC 22\,nm, 1.5\,GHz,
\texttt{compile\_ultra}). $\widehat{M}$~is the muxcount
defined in \S\ref{sec:fixed-cost}.}
\label{tab:fixed-decoder}
\smallskip
\begin{tabular}{l r r r}
\toprule
Layout & $\widehat{M}$ & Area ($\mu$m\textsuperscript{2}) & Power (mW) \\
\midrule
NV observed    & 1202 & 3360 & 2.81 \\
\S\ref{sec:fixed} synthesized
               &  121 & 2832 & 2.23 \\
\midrule
Reduction & $-$90\% & $-$16\% & $-$21\% \\
\bottomrule
\end{tabular}
\end{table}

Global slot assignment cuts extraction muxcount by 90\% relative to
the NVIDIA-observed 128-bit layout: $\widehat{M}$ falls from 1202 to
121 (solver lower bound 95), reducing total area by 16\% (3360 to
2832\,$\mu$m\textsuperscript{2}) and dynamic power by 21\% (2.81 to
2.23\,mW) under the same ASIC flow
(Table~\ref{tab:fixed-decoder}).
Both designs meet timing at 1.5\,GHz.
The NVIDIA-observed layout is a reconstructed baseline, not
necessarily optimized for $\widehat{M}$: production placement also
serves timing closure, datapath routing, and cross-generation
compatibility. The comparison measures the placement quality reached
under a decoder-extraction objective.
The synthesized layout is also more regular, with recurring field
names clustered at shared extraction positions.

\subsection{Instruction footprint and container distribution}
\label{sec:eval-footprint}

The footprint metric charges each instance by its synthesized
container width:
\[
  F = \sum_{i} \mathrm{container\_bits}(i)
      \;+\; 64 \cdot N_{\mathrm{MCSR\text{-}SET}}\,.
\]
The fixed 128-bit baseline is
$128 \times 1{,}107{,}272 = 141{,}730{,}816$~bits.
All reductions reported below include MCSR-SET overhead.

\begin{table}[t]
\centering
\caption{Variable-length encoding ablation. Reductions include
MCSR-SET overhead; area is generated decoder area.}
\label{tab:footprint-ablation}
\smallskip
\begin{tabular}{l r r r r r r}
\toprule
Variant & 64b & 96b & 128b & Red. & $\widehat{M}$ & Area \\
        & (\%) & (\%) & (\%) &      &               & ($\mu$m\textsuperscript{2}) \\
\midrule
\textit{pure}       & 32.3 & 58.2 &  9.5 & 30.7\% &  937 & 5054 \\
\textit{+singleton} & 34.5 & 59.3 &  6.1 & 32.1\% & 1044 & 4854 \\
\textit{+MCSR}      & 36.4 & 58.2 &  5.4 & 32.5\% & 1118 & 5046 \\
\textit{full}       & 38.3 & 56.6 &  5.1 & 33.1\% & 1227 & 4830 \\
\bottomrule
\end{tabular}
\end{table}

An ablation of the variable-length pipeline isolates each mechanism:
containers and prefix-free codes alone reduce footprint by 30.7\%
(128 to 89~bits average), and singleton absorption and MCSR offloading
contribute 1.4 and 1.8~pp over this baseline; the \textit{full} variant
combines both at 33.1\% (Table~\ref{tab:footprint-ablation}). MCSR offloading is optional: disabling it
leaves singleton absorption at 32.1\% with no added architectural
state. The \textit{full} variant adds 5{,}268 \texttt{MCSR-SET}
insertions ($\sim$0.5\% of the stream). Even a 4$\times$ larger
\texttt{MCSR-SET} count would keep the reduction above 32\%.

The footprint reduction is stable across architectures and suite
partitions. Full re-synthesis from each architecture's specification
reduces footprint by 29.6\% on Ampere and 33.1\% on both Hopper and
Blackwell (Fig.~\ref{fig:per-benchmark}); the corresponding medians
are 30.3\%, 33.2\%, and 32.4\%, so the cross-generation result is not
driven by a few large kernels. Blackwell keeps
141/142 inputs above 25\%. A leave-CUTLASS-out solve, removing the
suite that carries 90\% of instruction instances, drops aggregate
reduction only to 31.4\%, keeps held-out CUTLASS at 31.4\%, and changes
121/142 inputs by less than 2~pp. Coarse 64/96/128-bit classes are
constrained mainly by field-budget feasibility rather than fine-grained
frequency ranks. In the \textit{full} solution, 94.9\% of static
instances fit in 64- or 96-bit containers; only 5.1\% retain 128-bit
containers.

Generated decoder cost stays within
4.8--5.1k\,$\mu$m\textsuperscript{2} across variants and meets
1.5\,GHz timing; $\widehat{M}$ is not directly comparable because
singleton absorption and MCSR offloading change the decoded field set
(Table~\ref{tab:footprint-ablation}).

\begin{figure}[t]
\centering
\begin{subfigure}[b]{0.44\columnwidth}
  \centering
  \includegraphics[width=\columnwidth]{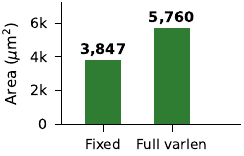}
  \caption{Generated logic}
  \label{fig:frontend-logic}
\end{subfigure}
\hfill
\begin{subfigure}[b]{0.54\columnwidth}
  \centering
  \includegraphics[width=\columnwidth]{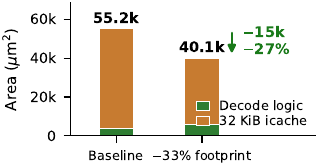}
  \caption{Per-subcore SRAM trade}
  \label{fig:frontend-trade}
\end{subfigure}
\caption{Generated frontend logic and same-node instruction-SRAM area
at TSMC 22\,nm. SRAM estimates use the TSMC 22\,nm memory compiler
and include bit-cell array area only.}
\label{fig:frontend-area}
\end{figure}

\subsection{Frontend logic and instruction-SRAM area}
\label{sec:eval-cost}

The added frontend logic is small relative to the instruction-SRAM it saves at the same technology node
(Fig.~\ref{fig:frontend-area}). The \textit{full} variable-length
frontend adds 1{,}913\,$\mu$m\textsuperscript{2} per subcore and still
meets 1.5\,GHz (Fig.~\ref{fig:frontend-logic}). Replicated across all 512
subcores of a GA100 configuration~\cite{nvidia-a100-whitepaper},
this is 0.98\,mm\textsuperscript{2}, or 0.12\% of the
826\,mm\textsuperscript{2} die
(0.006\% at 7\,nm under
DeepScaleTool~\cite{sarangi2021deepscaletool}).

For a same-node SRAM
reference, we generated 32\,KiB single-port instruction arrays with
the TSMC 22\,nm memory compiler; a 33\% smaller encoded stream needs
about 33\% less bit-cell capacity at equal byte-level headroom,
corresponding to about 17{,}000\,$\mu$m\textsuperscript{2} less
bit-cell area per subcore, roughly 9$\times$ the added decode logic
(Fig.~\ref{fig:frontend-trade}). Taken as capacity instead of area,
the same reduction stretches the instruction budgets of
\S\ref{sec:bg-supply} by about 1.5$\times$: a 128\,KiB array holds
roughly 12{,}000 variable-length instructions instead of 8{,}192.
Both readings are first-order:
they exclude tags and peripheral logic and do not redesign the
instruction-cache hierarchy. This trade-off becomes more favorable at
advanced nodes because SRAM density scales more slowly than logic
density~\cite{tsmc-n3-iedm}.

The variable-length frontend adds no pipeline stage and lowers fetch
pressure. Prefix inspection, container selection, and alignment sit
in the same combinational stage as field extraction: the
variable-length fetch-decode path closes at the same 0.67\,ns period
as the fixed-length path with 26 logic levels against 28, and both
designs keep one pipeline boundary at the decoded field vector, so
decode latency is one cycle in either encoding. On the supply side, a
Blackwell instruction occupies 85.4~bits of container on average on
this benchmark set (38.3\% 64-bit, 56.6\% 96-bit, and 5.1\% 128-bit),
or 85.7~bits once \texttt{MCSR-SET} insertions are charged, so one
128-bit L1 return carries 1.49 instructions where the fixed-length
stream carries 1.00. Decode logic is one component of the frontend,
and instruction-cache miss behavior, scheduler occupancy, and
end-to-end performance are outside this static and RTL-level
evaluation.

\section{Related Work and Discussion}
\label{sec:discussion}

\subsection{Encoding synthesis and field placement}
\label{sec:rw-encoding}

\begin{table}[t]
\centering
\caption{Prior work positioned by primary design variable and
evaluation artifact.}
\label{tab:rw-compare}
\footnotesize
\setlength{\tabcolsep}{3pt}
\begin{tabular*}{\columnwidth}{@{\extracolsep{\fill}}lll@{}}
\toprule
Approach & Design variable & Evaluation artifact \\
\midrule
Nohl et al.\ \cite{nohl2003lisa}  & per-GROUP opcode + fields & code density \\
PICO \cite{aditya1999pico}        & VLIW template layout  & compiler estimate \\
Lee et al.\ \cite{lee2007dutt}    & instruction selection  & area model \\
Chattopadhyay et al.\ \cite{chattopadhyay2007power}
                                   & opcode assignment      & power model \\
Thumb/RVC \cite{waterman2014riscvc}
                                   & compact subset         & ISA specification \\
Code compr.\ \cite{lefurgy1997codepress,wolfe1992coding}
                                   & dictionary encoding    & HW decompression \\
\textbf{This work}   & \textbf{cross-form field placement} & \textbf{footprint + ASIC} \\
\bottomrule
\end{tabular*}
\end{table}

Encoding synthesis from architecture description languages is a
long-standing line: Nohl et al.~\cite{nohl2003lisa} allocate
variable-length opcodes to minimize instruction word width
(synthesized widths are not byte-aligned), and Chattopadhyay et
al.~\cite{chattopadhyay2007power} reassign opcode values to
minimize instruction-bus switching power, both on LISA models of
register-dominated cores such as ARM7100 and MIPS32, and both
solved heuristically. A recent line instead redesigns ISA semantics
for density~\cite{maroun2025dense}. We cast SASS's cross-form
placement (\S\ref{sec:bg-motivation}) as one CP-SAT model over all
975 forms, and the solver reports a lower bound alongside the
layout (\S\ref{sec:eval-decoder}): the variable-length container
assignment (\S\ref{sec:varlen}) generalizes Nohl's variable-length
opcode to byte-aligned container classes, while the fixed-length
model (\S\ref{sec:fixed}) holds the opcode fixed and re-places
operand and modifier fields. Modern ADL toolchains (Synopsys ASIP
Designer~\cite{synopsys-asip-designer},
OpenVADL~\cite{huber2024vadl}) generate encodings without
documented placement algorithms; PICO~\cite{aditya1999pico}, Lee et
al.~\cite{lee2007dutt}, and Xiao et al.~\cite{xiao2026isamore}
optimize neighboring objects (VLIW templates; instruction
selection). Compact encodings are otherwise reached by
hand-designed formats (AMD multi-width
families~\cite{amd-cdna3-isa}; Thumb, MIPS16,
RVC~\cite{waterman2014riscvc}) or by dictionary compression below
the ISA boundary~\cite{lefurgy1997codepress,wolfe1992coding};
field-placement synthesis adds neither a new hand format nor a
decompression stage (Table~\ref{tab:rw-compare}).

\subsection{Applicability and limitations}
\label{sec:rw-discussion}

The CP-SAT model does not require a free-form layout: it can tie
field names together, pin selected fields to fixed slots, or leave
only selected operand families free. It applies when instruction
descriptions expose recurring semantic fields and the bit
layout is not frozen as a long-lived public machine-code contract.
This condition appears beyond SASS in GPU compiler stacks where
backend revisions absorb encoding changes: Mesa's Gen12/Xe bring-up
added new binary-encoding and disassembly paths for basic,
three-source, control-flow, SEND, compact, and datatype
instructions~\cite{mesa193gen12}, while Mali reverse-engineering
reports Bifrost-to-Valhall movement of dependency metadata from
clause headers into instruction
fields~\cite{collabora2021maliG78}. These cases show the freedom
to re-encode; applying the solver there would still require
recovering a per-form field specification as in
\S\ref{sec:foundations}. Where fields are few and regular,
placement has little to coordinate; the gains of
\S\ref{sec:eval} rest on SASS's modifier density
(\S\ref{sec:bg-motivation}).

Several limitations bound these results. The area trade-off in
\S\ref{sec:eval-cost} is a first-order comparison: we synthesize the
added frontend logic and compare it with memory-compiler SRAM arrays,
but do not redesign cache tags, banking, associativity, replacement,
or prefetch behavior. MCSR selection is based on observed modifier
change rates rather than compiler scheduling. The reported \texttt{MCSR-SET} count scans only
linear instruction order; control-flow-correct offloading adds
branch-merge and loop-entry resets that compiler dataflow would
refine, with the insertion overhead quantified in
\S\ref{sec:eval-footprint}. Compiler co-design that biases instruction
selection toward compressible forms is a natural extension.

\section{Conclusion}
\label{sec:conclusion}

Instruction layout can be treated as post-specification synthesis when
an ISA exposes recurring semantic fields whose bit placement is not
frozen by a small format catalogue. Starting from public SASS
artifacts, the validated specification and disassembler make this
structure precise enough for automated synthesis across four GPU
generations. On \texttt{sm\_100}, fixed-length slot assignment reduces
muxcount from 1202 to 121 and decoder area by 16\%; variable-length
synthesis reduces Blackwell footprint by 33\% with stable Ampere and
Hopper re-synthesis. The replicated frontend-area delta is 0.12\% of a
GA100-class die; at the subcore level, memory-compiler SRAM estimates
map this saving to about 9$\times$ the added frontend logic in
bit-cell area. These results position encoding synthesis as a
practical way to trade small decode-logic overhead for instruction
capacity headroom when the machine-code contract leaves field
placement available.

\section*{Acknowledgements}

This work was supported in part by the National Key Research and
Development Program of China under Grant 2022YFB4500101 and in part by
Shanghai Municipal Science and Technology Major Project.


\bibliographystyle{IEEEtran}
\bibliography{refs}

\end{document}